\documentclass[%
reprint,
superscriptaddress,
showpacs,
 amsmath,amssymb,
 aps
]{revtex4-1}

\usepackage{graphicx}
\usepackage{dcolumn}
\usepackage{bm}
\usepackage{amsmath}
\usepackage{autobreak}
\usepackage{hyperref}
\usepackage[mathlines]{lineno}
\usepackage{natbib}
\usepackage{multirow}
\usepackage{graphicx}
\usepackage{float}
\usepackage{subfigure}
\usepackage{ragged2e}
\usepackage[dvipsnames]{xcolor}
\usepackage[fontsize=10pt]{fontsize}
\usepackage{url}
\usepackage{makecell}
\begin{document}

\preprint{APS/123-QED}

\title{Projected WIMP sensitivity of the CDEX-50 dark matter experiment}

\author{X.~P.~Geng}
\affiliation{Key Laboratory of Particle and Radiation Imaging (Ministry of Education) and Department of Engineering Physics, Tsinghua University, Beijing 100084}
\author{L.~T.~Yang}
\email{Corresponding author: yanglt@mail.tsinghua.edu.cn}
\affiliation{Key Laboratory of Particle and Radiation Imaging (Ministry of Education) and Department of Engineering Physics, Tsinghua University, Beijing 100084}
\author{Q.~Yue}
\email{Corresponding author: yueq@mail.tsinghua.edu.cn}
\affiliation{Key Laboratory of Particle and Radiation Imaging (Ministry of Education) and Department of Engineering Physics, Tsinghua University, Beijing 100084}
\author{K.~J.~Kang}
\affiliation{Key Laboratory of Particle and Radiation Imaging (Ministry of Education) and Department of Engineering Physics, Tsinghua University, Beijing 100084}
\author{Y.~J.~Li}
\affiliation{Key Laboratory of Particle and Radiation Imaging (Ministry of Education) and Department of Engineering Physics, Tsinghua University, Beijing 100084}

\author{H.~P.~An}
\affiliation{Key Laboratory of Particle and Radiation Imaging (Ministry of Education) and Department of Engineering Physics, Tsinghua University, Beijing 100084}
\affiliation{Department of Physics, Tsinghua University, Beijing 100084}

\author{Greeshma~C.}
\altaffiliation{Participating as a member of TEXONO Collaboration}
\affiliation{Institute of Physics, Academia Sinica, Taipei 11529}

\author{J.~P.~Chang}
\affiliation{NUCTECH Company, Beijing 100084}

\author{Y.~H.~Chen}
\affiliation{YaLong River Hydropower Development Company, Chengdu 610051}
\author{J.~P.~Cheng}
\affiliation{Key Laboratory of Particle and Radiation Imaging (Ministry of Education) and Department of Engineering Physics, Tsinghua University, Beijing 100084}
\affiliation{College of Nuclear Science and Technology, Beijing Normal University, Beijing 100875}
\author{W.~H.~Dai}
\affiliation{Key Laboratory of Particle and Radiation Imaging (Ministry of Education) and Department of Engineering Physics, Tsinghua University, Beijing 100084}
\author{Z.~Deng}
\affiliation{Key Laboratory of Particle and Radiation Imaging (Ministry of Education) and Department of Engineering Physics, Tsinghua University, Beijing 100084}
\author{C.~H.~Fang}
\affiliation{College of Physics, Sichuan University, Chengdu 610065}

\author{H.~Gong}
\affiliation{Key Laboratory of Particle and Radiation Imaging (Ministry of Education) and Department of Engineering Physics, Tsinghua University, Beijing 100084}
\author{Q.~J.~Guo}
\affiliation{School of Physics, Peking University, Beijing 100871}
\author{T.~Guo}
\affiliation{Key Laboratory of Particle and Radiation Imaging (Ministry of Education) and Department of Engineering Physics, Tsinghua University, Beijing 100084}
\author{X.~Y.~Guo}
\affiliation{YaLong River Hydropower Development Company, Chengdu 610051}
\author{L.~He}
\affiliation{NUCTECH Company, Beijing 100084}
\author{S.~M.~He}
\affiliation{YaLong River Hydropower Development Company, Chengdu 610051}
\author{J.~W.~Hu}
\affiliation{Key Laboratory of Particle and Radiation Imaging (Ministry of Education) and Department of Engineering Physics, Tsinghua University, Beijing 100084}
\author{H.~X.~Huang}
\affiliation{Department of Nuclear Physics, China Institute of Atomic Energy, Beijing 102413}
\author{T.~C.~Huang}
\affiliation{Sino-French Institute of Nuclear and Technology, Sun Yat-sen University, Zhuhai 519082}
\author{L.~Jiang}
\affiliation{Key Laboratory of Particle and Radiation Imaging (Ministry of Education) and Department of Engineering Physics, Tsinghua University, Beijing 100084}
\author{S.~Karmakar}
\altaffiliation{Participating as a member of TEXONO Collaboration}
\affiliation{Institute of Physics, Academia Sinica, Taipei 11529}

\author{H.~B.~Li}
\altaffiliation{Participating as a member of TEXONO Collaboration}
\affiliation{Institute of Physics, Academia Sinica, Taipei 11529}
\author{H.~Y.~Li}
\affiliation{College of Physics, Sichuan University, Chengdu 610065}
\author{J.~M.~Li}
\affiliation{Key Laboratory of Particle and Radiation Imaging (Ministry of Education) and Department of Engineering Physics, Tsinghua University, Beijing 100084}
\author{J.~Li}
\affiliation{Key Laboratory of Particle and Radiation Imaging (Ministry of Education) and Department of Engineering Physics, Tsinghua University, Beijing 100084}
\author{Q.~Y.~Li}
\affiliation{College of Physics, Sichuan University, Chengdu 610065}
\author{R.~M.~J.~Li}
\affiliation{College of Physics, Sichuan University, Chengdu 610065}
\author{X.~Q.~Li}
\affiliation{School of Physics, Nankai University, Tianjin 300071}
\author{Y.~L.~Li}
\affiliation{Key Laboratory of Particle and Radiation Imaging (Ministry of Education) and Department of Engineering Physics, Tsinghua University, Beijing 100084}
\author{Y.~F.~Liang}
\affiliation{Key Laboratory of Particle and Radiation Imaging (Ministry of Education) and Department of Engineering Physics, Tsinghua University, Beijing 100084}
\author{B.~Liao}
\affiliation{College of Nuclear Science and Technology, Beijing Normal University, Beijing 100875}
\author{F.~K.~Lin}
\altaffiliation{Participating as a member of TEXONO Collaboration}
\affiliation{Institute of Physics, Academia Sinica, Taipei 11529}
\author{S.~T.~Lin}
\affiliation{College of Physics, Sichuan University, Chengdu 610065}
\author{J.~X.~Liu}
\affiliation{Key Laboratory of Particle and Radiation Imaging (Ministry of Education) and Department of Engineering Physics, Tsinghua University, Beijing 100084}
\author{S.~K.~Liu}
\affiliation{College of Physics, Sichuan University, Chengdu 610065}
\author{Y.~D.~Liu}
\affiliation{College of Nuclear Science and Technology, Beijing Normal University, Beijing 100875}
\author{Y.~Liu}
\affiliation{College of Physics, Sichuan University, Chengdu 610065}
\author{Y.~Y.~Liu}
\affiliation{College of Nuclear Science and Technology, Beijing Normal University, Beijing 100875}
\author{H.~Ma}
\affiliation{Key Laboratory of Particle and Radiation Imaging (Ministry of Education) and Department of Engineering Physics, Tsinghua University, Beijing 100084}
\author{Y.~C.~Mao}
\affiliation{School of Physics, Peking University, Beijing 100871}
\author{Q.~Y.~Nie}
\affiliation{Key Laboratory of Particle and Radiation Imaging (Ministry of Education) and Department of Engineering Physics, Tsinghua University, Beijing 100084}
\author{J.~H.~Ning}
\affiliation{YaLong River Hydropower Development Company, Chengdu 610051}
\author{H.~Pan}
\affiliation{NUCTECH Company, Beijing 100084}
\author{N.~C.~Qi}
\affiliation{YaLong River Hydropower Development Company, Chengdu 610051}
\author{J.~Ren}
\affiliation{Department of Nuclear Physics, China Institute of Atomic Energy, Beijing 102413}
\author{X.~C.~Ruan}
\affiliation{Department of Nuclear Physics, China Institute of Atomic Energy, Beijing 102413}
\author{M.~K.~Singh}
\altaffiliation{Participating as a member of TEXONO Collaboration}
\affiliation{Institute of Physics, Academia Sinica, Taipei 11529}
\affiliation{Department of Physics, Banaras Hindu University, Varanasi 221005}
\author{T.~X.~Sun}
\affiliation{College of Nuclear Science and Technology, Beijing Normal University, Beijing 100875}
\author{C.~J.~Tang}
\affiliation{College of Physics, Sichuan University, Chengdu 610065}

\author{Y.~Tian}
\affiliation{Key Laboratory of Particle and Radiation Imaging (Ministry of Education) and Department of Engineering Physics, Tsinghua University, Beijing 100084}
\author{G.~F.~Wang}
\affiliation{College of Nuclear Science and Technology, Beijing Normal University, Beijing 100875}
\author{J.~Z.~Wang}
\affiliation{Key Laboratory of Particle and Radiation Imaging (Ministry of Education) and Department of Engineering Physics, Tsinghua University, Beijing 100084}
\author{L.~Wang}
\affiliation{Department of  Physics, Beijing Normal University, Beijing 100875}
\author{Q.~Wang}
\affiliation{Key Laboratory of Particle and Radiation Imaging (Ministry of Education) and Department of Engineering Physics, Tsinghua University, Beijing 100084}
\affiliation{Department of Physics, Tsinghua University, Beijing 100084}
\author{Y.~F.~Wang}
\affiliation{Key Laboratory of Particle and Radiation Imaging (Ministry of Education) and Department of Engineering Physics, Tsinghua University, Beijing 100084}
\author{Y.~X.~Wang}
\affiliation{School of Physics, Peking University, Beijing 100871}
\author{H.~T.~Wong}
\altaffiliation{Participating as a member of TEXONO Collaboration}
\affiliation{Institute of Physics, Academia Sinica, Taipei 11529}
\author{S.~Y.~Wu}
\affiliation{YaLong River Hydropower Development Company, Chengdu 610051}
\author{Y.~C.~Wu}
\affiliation{Key Laboratory of Particle and Radiation Imaging (Ministry of Education) and Department of Engineering Physics, Tsinghua University, Beijing 100084}
\author{H.~Y.~Xing}
\affiliation{College of Physics, Sichuan University, Chengdu 610065}
\author{R. Xu}
\affiliation{Key Laboratory of Particle and Radiation Imaging (Ministry of Education) and Department of Engineering Physics, Tsinghua University, Beijing 100084}
\author{Y.~Xu}
\affiliation{School of Physics, Nankai University, Tianjin 300071}
\author{T.~Xue}
\affiliation{Key Laboratory of Particle and Radiation Imaging (Ministry of Education) and Department of Engineering Physics, Tsinghua University, Beijing 100084}
\author{Y.~L.~Yan}
\affiliation{College of Physics, Sichuan University, Chengdu 610065}

\author{N.~Yi}
\affiliation{Key Laboratory of Particle and Radiation Imaging (Ministry of Education) and Department of Engineering Physics, Tsinghua University, Beijing 100084}
\author{C.~X.~Yu}
\affiliation{School of Physics, Nankai University, Tianjin 300071}
\author{H.~J.~Yu}
\affiliation{NUCTECH Company, Beijing 100084}
\author{J.~F.~Yue}
\affiliation{YaLong River Hydropower Development Company, Chengdu 610051}
\author{M.~Zeng}
\affiliation{Key Laboratory of Particle and Radiation Imaging (Ministry of Education) and Department of Engineering Physics, Tsinghua University, Beijing 100084}
\author{Z.~Zeng}
\affiliation{Key Laboratory of Particle and Radiation Imaging (Ministry of Education) and Department of Engineering Physics, Tsinghua University, Beijing 100084}
\author{B.~T.~Zhang}
\affiliation{Key Laboratory of Particle and Radiation Imaging (Ministry of Education) and Department of Engineering Physics, Tsinghua University, Beijing 100084}
\author{F.~S.~Zhang}
\affiliation{College of Nuclear Science and Technology, Beijing Normal University, Beijing 100875}
\author{L.~Zhang}
\affiliation{College of Physics, Sichuan University, Chengdu 610065}
\author{Z.~H.~Zhang}
\affiliation{Key Laboratory of Particle and Radiation Imaging (Ministry of Education) and Department of Engineering Physics, Tsinghua University, Beijing 100084}
\author{Z.~Y.~Zhang}
\affiliation{Key Laboratory of Particle and Radiation Imaging (Ministry of Education) and Department of Engineering Physics, Tsinghua University, Beijing 100084}
\author{J.~Z.~Zhao}
\affiliation{Key Laboratory of Particle and Radiation Imaging (Ministry of Education) and Department of Engineering Physics, Tsinghua University, Beijing 100084}
\author{K.~K.~Zhao}
\affiliation{College of Physics, Sichuan University, Chengdu 610065}
\author{M.~G.~Zhao}
\affiliation{School of Physics, Nankai University, Tianjin 300071}
\author{J.~F.~Zhou}
\affiliation{YaLong River Hydropower Development Company, Chengdu 610051}
\author{Z.~Y.~Zhou}
\affiliation{Department of Nuclear Physics, China Institute of Atomic Energy, Beijing 102413}
\author{J.~J.~Zhu}
\affiliation{College of Physics, Sichuan University, Chengdu 610065}

\collaboration{CDEX Collaboration}
\noaffiliation

\date{\today}

\begin{abstract}
CDEX-50 is a next-generation project of the China Dark Matter Experiment (CDEX) that aims to search for dark matter using a 50-kg germanium detector array. This paper comprises a thorough summary of the CDEX-50 dark matter experiment, including an investigation of potential background sources and the development of a background model. Based on the baseline model, the projected sensitivity of weakly interacting massive particle (WIMP) is also presented. The expected background level within the energy region of interest, set to 2--2.5 keVee, is $\sim$0.01 counts keVee$^{-1}$ kg$^{-1}$ day$^{-1}$. At 90\% confidence level, the expected sensitivity to spin-independent WIMP-nucleon couplings is estimated to reach a cross-section of 5.1 $\times$ 10$^{-45}$ cm$^{2}$ for a WIMP mass of 5 GeV/c$^{2}$ with an exposure objective of 150 kg$\cdot$year and an analysis threshold of 160 eVee. This science goal will correspond to the most sensitive results for WIMPs with a mass of 2.2--8 GeV/c$^{2}$.
\end{abstract}

\maketitle

\section{Introduction}\label{sec:introduction}
The existence of dark matter (DM), as indicated by diverse astrophysical and cosmological observations at different scales~\cite{DM}, has been one of the most significant problems in physics for a long time. Weakly interacting massive particles (WIMPs, denoted as $\chi$) have emerged the most promising candidate for DM, and have been extensively explored through various direct detection experiments for decades~\cite{EDELWEISS,LUX,PICO,PandaX-4T,cdex1,cdex10,cdms,xenon,cogent,darkside,LZ,CRESST,SENSEI,DAMIC}. 

Based on $p$-type point contact germanium ($p$PCGe) detectors, which offer advantages in energy resolution and threshold~\cite{soma2016}, the China Dark Matter Experiment (CDEX) has been dedicated to DM direct detection experiments. These experiments have been conducted at the China Jinping Underground Laboratory (CJPL)~\cite{CJPL} for many years and have spanned two phases. The first phase of the experiment, CDEX-1~\cite{cdex1}, was installed and started operation in 2010 using a 1-kg single element $p$PCGe detector cooled by a cold-finger system. In 2016, the experiment was upgraded to the second phase, CDEX-10~\cite{cdex10}. This phase involved immersing a 10-kg $p$PCGe array directly in liquid nitrogen ($\rm{LN}_2$) for cooling. The detector array comprises three triple-element $p$PCGe detector strings encapsulated within a vacuum cryostat. Both phases have achieved world-leading results in the direct detection of DM and related research areas~\cite{cdex1,cdex12014,cdex1b2018,cdex1b_am,cdex10,cdex10_tech,cdex_darkpthoton,mec1b,cdex10_eft,cjpless,CRDM,cdex_er,cdex10_exotic,cdex_neutrino,cdex10_PBHDM}. 

For the next-generation of the experiment, CDEX-50, an upgraded detector array consisting of 50 germanium detectors with a target mass of 50 kg will be deployed in a 1725 m$^{3}$ tank filled with LN$_{2}$. This tank is situated at Hall C1 as part of the extension project of CJPL (CJPL-II)~\cite{CJPL}. Improved purity of detector components, stringently controlled germanium exposure, and LN$_{2}$ tank shielding are expected to greatly reduce the background level. The expected analysis threshold is 160 eVee (``eVee'' represents electron equivalent energy derived from energy calibration), and the exposure goal is 150 kg$\cdot$year.

The remainder of the paper is organized as follows. The details of CDEX-50 detectors are described in section~\ref{sec:2}. The background source analysis, corresponding simulation, and the background model are presented in section~\ref{sec:3}. In section~\ref{sec:4}, the projected sensitivity on WIMP is derived based on the background model.

\section{CDEX-50 experiment}~\label{sec:2}
CDEX-50 will be deployed as an array comprising 5 strings, each consisting of 10 detectors. This array will be directly immersed into a tank filled with LN$_{2}$. The back-end data-acquisition system will assign coincident triggers to events because WIMP and other exotic particles are expected to interact with individual detectors over a period of time.

\subsection{CDEX-50 detector unit}~\label{sec:details.1}
\begin{figure}[htbp]
	\centering\includegraphics[width=\columnwidth]{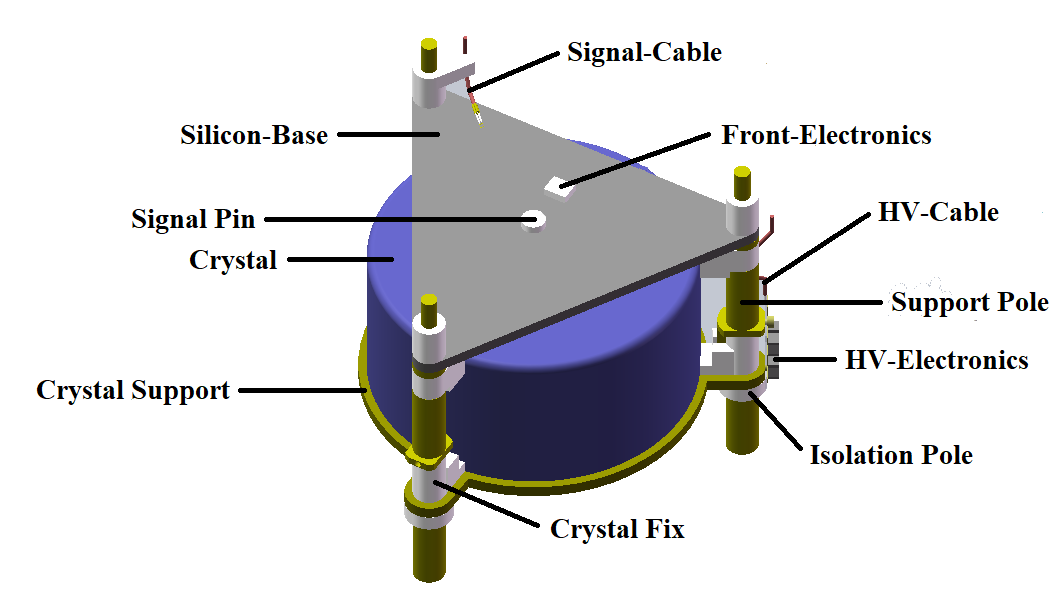}
	\caption{Render of the CDEX-50 detector unit. Different colors correspond to different materials. The germanium crystal is purple, and the Signal Pin, Crystal Support, and Support Pole (all made from copper) are yellow. Components such as Crystal Fix and Isolation Pole, constructed from PTFE , are represented in white. The Silicon-Base is illustrated in gray.}
	\label{fig:unit}
\end{figure}

The germanium detector unit comprises several components: the germanium crystal, supporting structure, and electronics with high voltage (HV) and signal cables. The germanium crystal is designed to be a cylinder with a diameter of 80 mm, length of 40 mm, and mass of $\sim$1 kg. An inactive layer thickness of 1.0 mm is estimated based on previous analyses of CDEX germanium detectors~\cite{deadlayer1,deadlayer2}. The supporting structure bears the weight of the detector and can fix the crystal and electronics in the event of disturbance in LN$_2$ during deployment and operation. The electronics are used to supply the appropriate HV for the crystal and collect signals from the interactions in the volume, accomplished through a signal pin that links the front-end electronic to the electrode of the crystal. The design diagram of the detector unit is shown in figure~\ref{fig:unit}, rendered by Geant4~\cite{geant43}. The components are designed with low mass to achieve ultralow radioactivity while maintaining the expected functionality. A comprehensive list of components of the CDEX-50 detector is presented in table~\ref{tab:unit}.

\renewcommand{\arraystretch}{1.3}
\begin{table}[!htbp]
	\begin{ruledtabular}
	\caption{Components of the CDEX-50 detector unit. They are classified based on their functionality.}
	\label{tab:unit}
	\centering
	\begin{tabular}{cccc} 
		Category & Component & Material & Mass [g]\\
		\hline
		Crystal&Crystal&Germanium (Ge)&$\sim$1000\\
		\hline
		\multirow{2}{*}{Cabling}
		&HV-Cable &PTFE, Copper& 0.39 \\
		&Signal-Cable &PTFE, Copper& 0.33\\
		\hline
		\multirow{2.5}{*}{Electronics}
		&Signal Pin&Copper&0.01\\
		&HV/Front-Electronics&\thead{PCB\\Resistor, Capacitor}&0.7\\
		
		\hline
		\multirow{5}{*}{Support}
		&Crystal Support&Copper&27.6\\
		&Support Pole &Copper&58.1\\
		&Crystal Fix&PTFE&17.8\\
		&Isolation Pole&PTFE&8.4\\
		&Silicon-Base &Silicon&30.2\\
		
		\hline
		TOTAL &
		&&1143.53\\
	\end{tabular}
	\end{ruledtabular}
\end{table}

\subsection{CDEX-50 detector array}\label{sec:2.2}
\begin{figure}[htbp]
	\centering
	\includegraphics[width=0.4\linewidth]{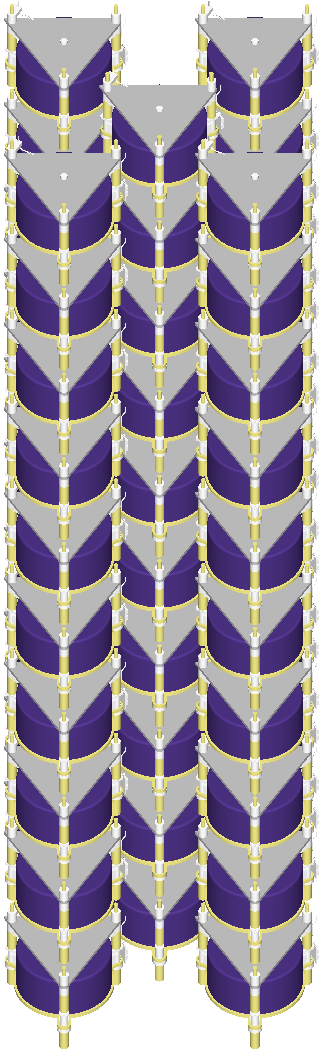}
	\caption{Render of the CDEX-50 detector array with a diameter and height of 348 mm and 938 mm, respectively.}
	\label{fig:array}
\end{figure}

The CDEX-50 array consists of 5 strings, each consisting of 10 detector units. The distance between two adjacent crystals within a string is 54 mm. Within each string, the detector units are supported by a clean material and the cables are carefully arranged to facilitate connection to back-end electronics outside the tank. The distance between two adjacent strings is 40 mm. During the deployment, the array will be positioned at the center of the tank, with all detector strings oriented vertically to the ground. They are deployed circularly, as shown in figure~\ref{fig:array}, rendered by Geant4~\cite{geant43}. 

\subsection{Cryostat and shielding}~\label{sec:2.3}
A large LN${_2}$ cryostat serves as the cryogenic system and shield against environmental radiation. The shape of the LN${_2}$ inside the cryostat nearly represents a cylinder with a diameter of 13 m and length of 13 m. The CDEX-50 detector array will be deployed at the center of the tank, and LN$_{2}$ of 6.5 m can shield the array from any angle (See Figure 6$a$ in Ref.~\cite{CJPL}).

\section{CDEX-50 backgrounds}~\label{sec:3}
Signals from background sources may be indistinguishable from those generated by WIMPs or other exotic particles. As a result, a comprehensive analysis of the background sources of CDEX-50 has been conducted to fully understand its background characteristics. This analysis encompasses both cosmogenic and primordial radionuclides arising from the environment and detector components as well as contributions from solar neutrinos. The specific activities of the radionuclides in different entities are measured or estimated, and their expected contribution is evaluated through Monte Carlo (MC) simulation. The background arising from solar neutrinos is estimated within the standard model (SM)~\cite{solarneutrino,coherent}. Furthermore, the contribution of each source in the energy region of interest (ROI), which is set to 2--2.5 keVee, is estimated. Based on the analysis, the background model is built. In this work, the energy resolution of CDEX-10~\cite{cdex10_tech} is adopted, which is characterized by the standard deviation of $\rm 35.8 + 16.6\times E^{\frac{1}{2}}(eV)$, where $\rm E$ is in keV.

\subsection{Background sources from environment}~\label{sec:LN2}
The main environmental concerns in underground laboratory are muon-induced, neutron, and $\gamma$ backgrounds. The muon-induced background is estimated to be $\textless$ 1.09$\times$10$^{-6}$ counts keVee$^{-1}$ kg$^{-1}$ day$^{-1}$ (cpkkd) in 0--4 keVee, this minimal contribution is due to the negligible muon flux at CJPL~\cite{muonflux}. Neutron-related effects in CJPL are predominantly attributed to spontaneous fission from $^{238}$U and ($\alpha$,n) reactions, where the $\alpha$ particles originate from uranium and thorium series radionuclides, the related radionuclides are mainly present in the wall of the laboratory~\cite{neutron}. $\gamma$ background mainly arises from decays of long-lived radionuclides in rocks and concrete~\cite{cjplII}. To mitigate these concerns, a 6.5-m thick LN$_{2}$ shield is employed to lower the environmental neutron and $\gamma$ background, which are estimated to be $\textless$ 3.84$\times$10$^{-9}$ and $\textless$ 3.21$\times$10$^{-6}$ cpkkd in 0--4 keVee, respectively~\cite{neutron,HUQINGDONG,cjplII,SHEZE}. These are negligible levels compared to the contribution from other background sources. Moreover, an unavoidable presence of $^{222}$Rn in LN${_2}$ leads to the emission of moderately high-energy $\gamma$-rays through the decay of $^{214}$Pb and $^{214}$Bi progeny. These decays occurring around the detector can contribute to the background. The specific activity of $^{222}$Rn after purification is expected to be 0.4 $\mu$Bq/kg~\cite{LN2}.

\subsection{Background sources from detector components}
The cosmogenic and primordial radionuclides are considered for the detector components, which are evaluated via MC simulation and measurements, respectively.

\subsubsection{Cosmogenic radioactivity}
Cosmogenic radionuclides produced via cosmic-ray activation in germanium crystal and copper are evaluated via MC simulation. CRY library~\cite{cry} is applied to generate spectra of cosmic-rays including neutron, proton, muon, and $\gamma$-ray, whereas Geant4~\cite{geant43} with $Shielding$ physics list is used to simulate particle interactions between cosmic-ray and crystal$\&$copper. The specific activities are derived according to the expected manufacturing/processing steps of crystal$\&$copper for CDEX-50~\cite{geactivation,copperactivation}.

The manufacturing/processing steps of germanium crystal and detector of CDEX-50 are controlled strictly to reduce the surface exposure time, and all the procedures have been optimized to reduce the time cost. During fabrication, the crystal is temporarily stored underground with an overburden of 50 m of water-equivalent (m.w.e.) adjacent to the worksite when not being processed, and the relocation is daily. The transportation is at low altitudes to avoid high cosmic-ray fluxes. During transportation, the crystal will be shielded by a low-carbon steel shield with 65 cm above and 36 cm on the sides. After arriving at the CJPL, the crystal will be stored for $\sim$3 years, which is the expected cooling time, before preparing for the physical operation. During this time, the specific activities of cosmogenic radionuclides will decrease through decay. The manufacturing/processing steps and specific activities of the considered radionuclides are presented in tables~\ref{tab:process} and ~\ref{tab:cos}, respectively, compared with a competitive Ge-based experiment, the MAJORANA DEMONSTRATOR (denoted as MJD)~\cite{MJD,MJD2,MJD3,MJD4,MJD5}, in cosmogenic radionuclide-induced background sensitivity. The $^{3}$H radioactivity of CDEX-50, which contributes most in the low-energy region, is lower than that of MJD by a factor of $\sim$6. This difference is due to the more time-consuming unshielded fabrication step employed by MJD and the daily relocation to the underground site utilized by CDEX-50.  While the cooling time of MJD is 1 year longer than that of CDEX-50,$^{3}$H can barely be reduced via cooling down of 1 year for a 12.32-year half-life, and the extended duration of unshielded conditions has a negligible impact due to the effective transportation shielding of CDEX-50. The difference in radioactivity of $^{68}$Ge, $^{65}$Zn, and $^{55}$Fe is less pronounced. This is because $^{68}$Ge and $^{65}$Zn with short half-life can be efficiently removed through cooling down, and the production rate of $^{55}$Fe is lower than that of $^{3}$H by a factor of $\sim$10.

9 radionuclides in germanium are considered for relatively long half-lives. Their expected specific activities after 3 years of cooling down evaluated through MC simulation are presented in table~\ref{tab:cos}.
\renewcommand{\arraystretch}{1.3}
\begin{table}[!htbp]
	\begin{ruledtabular}
	\caption{The manufacturing/processing steps of the germanium crystal and detector~\cite{geactivation} of CDEX-50, along with those of MJD~\cite{MJD,MJD2,MJD3,MJD4,MJD5}. The unit duration of each manufacturing/processing step is day. Note that the ``Underground storage'' of CDEX-50 is performed daily.}
	\label{tab:process}
	\centering
	\begin{tabular}{cccc}
		Manufacturing/\\processing step & CDEX-50 & MJD & Shielding\\
		\hline
		Fabrication & 60&$\sim$110&Underground storage\\
		Transportation& 65&$\sim$7&Transportation shield\\
		Cooling time&$\sim$1095 &$\sim$1460&Underground laboratory\\
	\end{tabular}
	\end{ruledtabular}
\end{table}
\renewcommand{\arraystretch}{1.3}
\begin{table}[!htbp]
	\begin{ruledtabular}
	\caption{Specific activities of the cosmogenic radionuclides of the germanium crystal after 3 years of cooling down, along with MJD with approximately 4 years of cooling down derived from best fit based on the observed spectrum~\cite{MJD2,MJD4}. The specific activity of $^{68}$Ga is considered the same as $^{68}$Ge because the progeny of $^{68}$Ge, $^{68}$Ga has a much shorter half life compared to $^{68}$Ge.}
	\label{tab:cos}
	\centering
	\begin{tabular}{ccc}
		Radionuclide & CDEX-50 [$\mu$Bq/kg]&MJD [$\mu$Bq/kg]\\
		\hline
		$^{3}$H&9.93E-01&(5.82$\pm$0.36)E+00\\	
		$^{49}$V&4.11E-02&NA\\	
		$^{54}$Mn&2.88E-02&NA\\
		$^{55}$Fe&1.66E-01&(1.48$\pm$2.20)E-01\\
		$^{57}$Co&6.96E-02&NA\\
		$^{60}$Co&6.64E-02&NA\\
		$^{63}$Ni&1.80E-02&NA\\
		$^{65}$Zn&3.26E-01&(2.08$\pm$2.12)E-01\\
		$^{68}$Ge&1.06E+00&(2.64$\pm$2.09)E-01\\		
	\end{tabular}
	\end{ruledtabular}
\end{table}

After production, the copper will be transported to CJPL and stored in CJPL with the crystal. 3 radionuclides in copper are considered for relatively long half-lives, and their specific activities after 3 years of cooling are presented in table~\ref{tab:coscopper}.
\renewcommand{\arraystretch}{1.3}
\begin{table}[!htbp]
	\begin{ruledtabular}
	\caption{Specific activities of the cosmogenic radionuclides of copper after 3 years of cooling down~\cite{copperactivation}.}
	\label{tab:coscopper}
	\centering
	\begin{tabular}{cc}
		Radionuclide & Specific activity [$\mu$Bq/kg]\\
		\hline
		$^{54}$Mn&2.38E-02\\	
		$^{57}$Co&5.32E-03\\
		$^{60}$Co&3.27E+00\\	
	\end{tabular}
	\end{ruledtabular}
\end{table}

\subsubsection{Ambient radioactivity}
Primordial radionuclides with long half-lives, including $^{238}$U and $^{232}$Th chains and $^{40}$K isotope, are introduced in the material of detector components unavoidably during manufacturing. Their specific activities can be measured using various methods, such as $\gamma$-ray spectrometry~\cite{GeTHU} and inductively coupled plasma-mass spectrometry (ICP-MS)~\cite{ICPMS}. All materials used in the detector components are selected by radioassay and presented in table~\ref{tab:pri}. 
\renewcommand{\arraystretch}{1.3}
\begin{table}[!htbp]
	\begin{ruledtabular}
	\caption{Specific activities ($\mu$Bq/kg) of the primordial radionuclides considered in the detector components. The data on copper comes from the measurement report of the 7N purity copper samples used in CDEX-50. Upper limits are given at 90\% confidence level (CL). The radionuclides are assumed to be distributed evenly in the material.}
	\label{tab:pri}
	\tabcolsep 5pt
	\centering
	\begin{tabular}{cccc}
		Components&$^{238}$U &$^{232}$Th &$^{40}$K \\
		\hline
		HV-Cable~\cite{SHEZE}&$<$2.4E+02&$<$1.2E+01&$<$3.2E+03\\
		Signal-Cable~\cite{SHEZE}&$<$5.1E+01&$<$1.9E+01&$<$2.9E+03\\
		Copper&$<$1.3E+00&$<$5.8E-01&$<$5.8E+00\\
		Electronics~\cite{legend,database,XENON_radioassay}&$<$4.1E+01&$<$2.2E+01&$<$3.1E+03\\
		PTFE~\cite{legend,database,XENON_radioassay}&$<$1.0E-01&$<$5.0E+00&$<$3.4E+02\\
		Silicon~\cite{SHEZE}&$<$3.8E+01&$<$1.9E+01&$<$3.0E+03\\	
	\end{tabular}
	\end{ruledtabular}
\end{table}

\subsection{Monte Carlo simulation of radionuclides}
The background simulation of radionuclides is conducted along with the Simulation and Analysis for Germanium Experiments (SAGE) package~\cite{SAGE}, a Geant4~\cite{geant43} application, with the integration of CDEX-50 geometry. The specific activities of the radionuclides arise from the discussion above. Each radionuclide in each component is simulated separately by generating 10$^{9}$ corresponding radionuclides. All the radioactive progenies that originated from $^{238}$U, $^{232}$Th, $^{222}$Rn, and $^{68}$Ge are considered, and decay chains are assumed to be in secular equilibrium. The location, deposited energy, and channel of every interaction for each event are recorded. Conversion from simulation result to background model is performed using the following formula:
\begin{equation}\label{eq_1}
	R\left[\frac{counts}{kg\cdot keV\cdot d}\right]=A\left[\frac{Bq}{kg}\right]\times M\left[{kg}\right]\times\frac{counts}{primaries}\times units,
\end{equation}
where $A$ denotes the specific activity of the radionuclides, $M$ denotes the total mass of specific components, $primaries$ is the number of simulated events, $counts$ is the number of events that deposit energy in the detector, and $units$ denotes calculation units involved.

Self-anticoincidence among the array is applied, and events with two or more trigger detectors during the detector response time (10 $\mu s$ for the germanium detector in this work) are tagged as coincidence. However, the cascade radiations from a single decay occurring within the active volume will be observed as single events. This is due to the brief time interval between radiations, which is considerably shorter than the detector response time. The survival probability of each radionuclide in each component is presented in table~\ref{tab:survival}. 

\renewcommand{\arraystretch}{1.3}
\begin{table}[!htbp]
	\begin{ruledtabular}
	\caption{Survival probability in ROI of each radionuclide in each component.}
	\label{tab:survival}
	\centering
		\resizebox{.98\columnwidth}{!}{
		\begin{tabular}{cccc}
			Radionuclide & Component & Survival Probability [\%]\\
			\hline
			$^{3}$H &Crystal&100.0\\
			\hline	
			$^{49}$V &Crystal&100.0\\
			\hline
			\multirow{4}{*}{$^{54}$Mn}
			&Crystal&82.57\\
			&Signal Pin&76.32\\
			&Crystal Support&71.22\\
			&Support Pole&80.55\\
			\hline
			$^{55}$Fe &Crystal&100.0\\
			\hline
			\multirow{4}{*}{$^{57}$Co}
			&Crystal&89.20\\
			&Signal Pin&71.17\\
			&Crystal Support&79.87\\
			&Support Pole&82.31\\
			\hline
			\multirow{4}{*}{$^{60}$Co}
			&Crystal&64.62\\
			&Signal Pin&64.22\\
			&Crystal Support&68.23\\
			&Support Pole&71.01\\
			
			\hline
			$^{63}$Ni &Crystal&100.0\\
			\hline
			$^{65}$Zn &Crystal&93.91\\
			\hline
			$^{68}$Ge &Crystal&100.0\\
			\hline
			$^{68}$Ga &Crystal&78.32\\
			\hline
			$^{222}$Rn &LN$_{2}$&73.64\\
			\hline
			\multirow{10}{*}{$^{238}$U}
			&HV-Cable&70.93\\
			&Signal-Cable&74.31\\
			&Signal Pin&75.80\\
			&HV-Electronics&73.17\\
			&Front-Electronics&73.40\\
			&Crystal Support&71.49\\
			&Support Pole&74.73\\
			&Crystal Fix&73.39\\
			&Isolation Pole&70.81\\
			&Silicon-Base&71.03\\
			
			\hline
			\multirow{10}{*}{$^{232}$Th}
			&HV-Cable&70.19\\
			&Signal-Cable&73.97\\			
			&Signal Pin&74.03\\
			&HV-Electronics&71.65\\
			&Front-Electronics&72.32\\
			&Crystal Support&74.23\\
			&Support Pole&75.52\\
			&Crystal Fix&71.36\\
			&Isolation Pole&70.89\\
			&Silicon-Base&71.27\\
			
			\hline
			\multirow{10}{*}{$^{40}$K}
			&HV-Cable&72.10\\
			&Signal-Cable&73.88\\
			&Signal Pin&74.49\\
			&HV-Electronics&77.15\\
			&Front-Electronics&78.89\\
			&Crystal Support&75.54\\
			&Support Pole&72.15\\
			&Crystal Fix&77.32\\
			&Isolation Pole&71.32\\
			&Silicon-Base&71.23\\				
		\end{tabular}
	}
	\end{ruledtabular}
\end{table}

The efficiencies of the self-anticoincidence are dependent on how the radionuclides deposit energy in the detector volume, determined by the emission and location of the radionuclides. The electron released by the decay of $^{3}$H within a crystal can barely reach other crystals for the maximum energy of 18.6 keV. This energy level is insufficient for penetration through LN$_{2}$ between crystals. Similarly, the decay radiations of $^{49}$V, $^{55}$Fe, $^{68}$Ge, and $^{63}$Ni also possess survival probabilities of $\sim$100\%. Among other cosmogenic radionuclides, some may deposit energy in ROI through low-energy characteristic x-rays, with relatively high-energy simultaneous $\gamma$ emission, which events can deposit energy across different crystals and get discarded for coincidence. This pattern also applies to primordial radionuclides with high-energy $\gamma$ emission, although they mainly deposit energy in ROI through the Compton effect for a relatively long gap filled with LN$_{2}$ from crystal.

\subsection{Background from solar neutrino}~\label{sec:cevns}
\begin{figure}[!htbp]
	\includegraphics[width=\columnwidth]{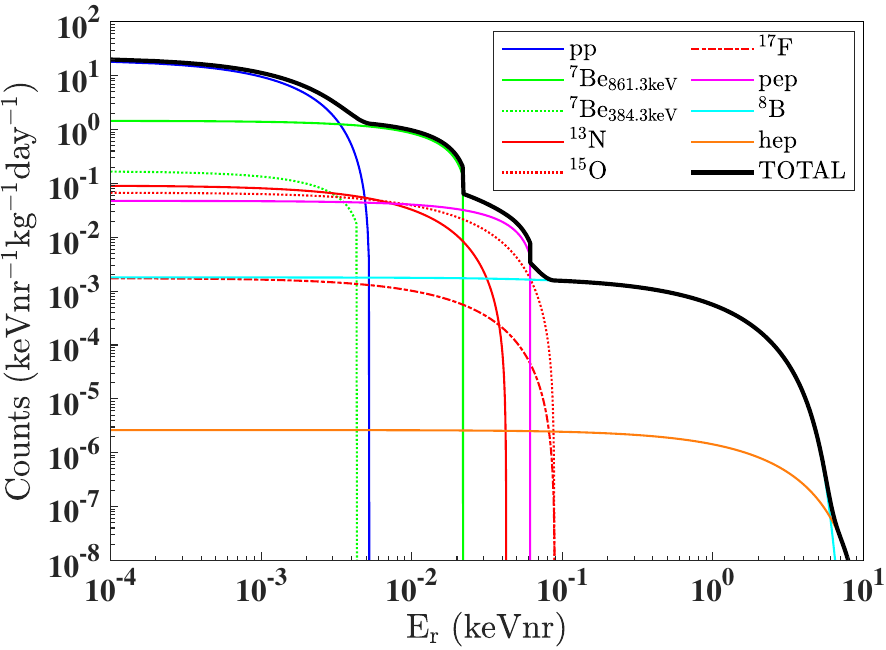}
    \includegraphics[width=\columnwidth]{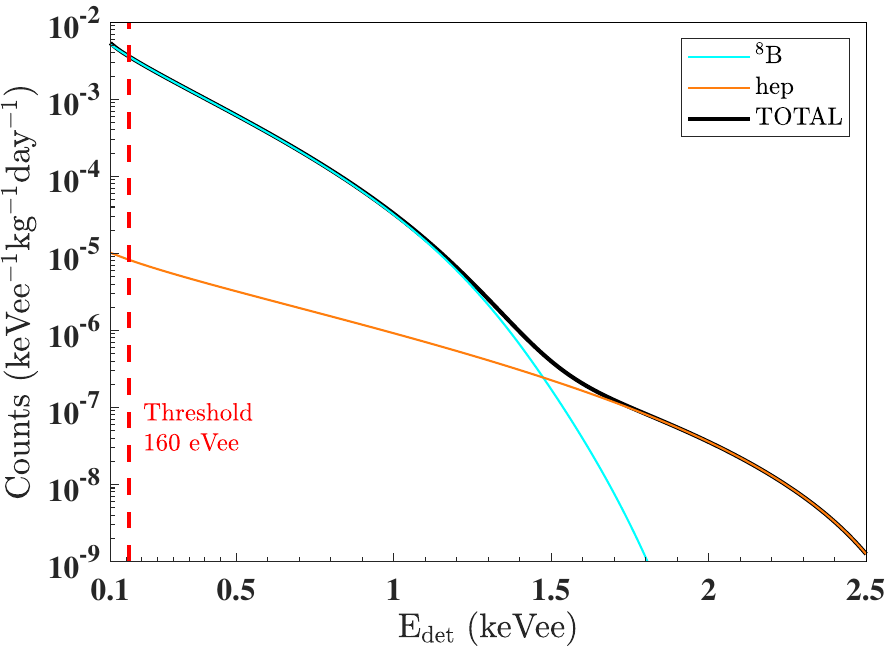}
	\caption{\textbf{Top:} Expected spectra of solar neutrino-nucleus scattering in Ge broken down into processes. \textbf{Bottom:} Expected spectra of deposited energy in CDEX-50 detectors with the consideration of quenching factor and energy resolution. The contribution of CE$\nu$NS is from $\rm ^{8}$B and $\rm hep$ neutrino considering a 160 eVee threshold.}
	\label{fig:cevns}
\end{figure}

Solar neutrinos may interact with both electron and nuclei in Ge through scattering and produce low-energy signals within the SM. Considering the typical $\mathcal{O}(100)~{\rm eV}$ energy threshold of the germanium detector, the contribution within the SM mainly comes from the coherent elastic neutrino-nucleus scattering (CE$\nu$NS). The differential cross-section and scattering rate of CE$\nu$NS are given as follows: 
\begin{align}
	\label{equ:crs}
	\frac{d\sigma(E_{r},E_{\nu})}{dE_r} &= \frac{G_{f}^{2}}{4\pi}Q_{w}^{2}(1-\frac{m_{N}E_{r}}{2E_{\nu}^{2}})F^{2}(E_{r}),\\
	\label{equ:scs}
	\frac{dR}{dE_{r}} &= N_{T}\int_{E_{\nu}^{min}}^{\infty}\frac{d\Phi}{dE_{\nu}}\frac{d\sigma(E_{r},E_{\nu})}{dE_r}dE_{\nu},
\end{align}
where $E_r$ is the recoil energy of the target, $E_\nu$ is the neutrino energy, $G_{f}$ is the Fermi constant, $Q_{w}$ is the weak nuclear charge, $m_{N}$ is the mass of the target nucleus, $F(E_{r})$ is the form factor, which is the Helm form factor~\cite{form,formfactor} adopted in this work, $N_{T}$ is the number of target nuclei per unit of mass of the detector material, $E_{\nu}^{min}$ is the minimum neutrino energy required to generate recoil energy $E_r$, and $\frac{d\Phi}{dE_{\nu}}$ is the differential flux of neutrinos. The B16-GS98 solar model (also referred to as the high-metallicity or HZ model) is adopted, and the values for the solar neutrino fluxes are taken from Ref.~\cite{solar_neutrino_flux}. The expected event rates in germanium detectors from solar neutrino-nucleus scattering and corresponding spectra in CDEX-50 detectors are shown in figure~\ref{fig:cevns}. The deposited energy $E_{\rm det}$ is modified by $E_{\rm det}=Q_{\rm nr}E_r$, where the quenching factor $Q_{\rm nr}$ in Ge is obtained using the TRIM package~\cite{TRIM}. 

The contribution from solar neutrino in ROI is 1.32$\times$10$^{-8}$ cpkkd, mainly from $\rm ^{8}B$ and $\rm hep$ neutrino-nucleus scattering. However, the effect of CE$\nu$NS cannot be exhibited through the contribution in ROI because the CE$\nu$NS background drops steeply below 2.5 keVee, and it is more appropriate to estimate the effect through the contribution of 0.16--0.5 keVee, which is 1.61$\times$10$^{-3}$ cpkkd.

\begin{figure}[!tbp]
	\centering\includegraphics[width=\columnwidth]{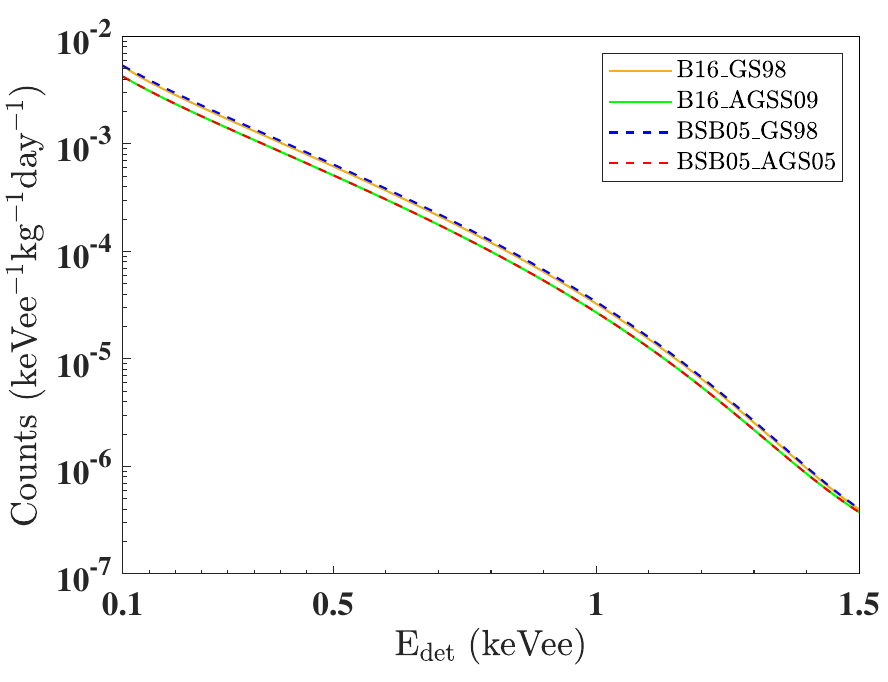}
	\centering\includegraphics[width=\columnwidth]{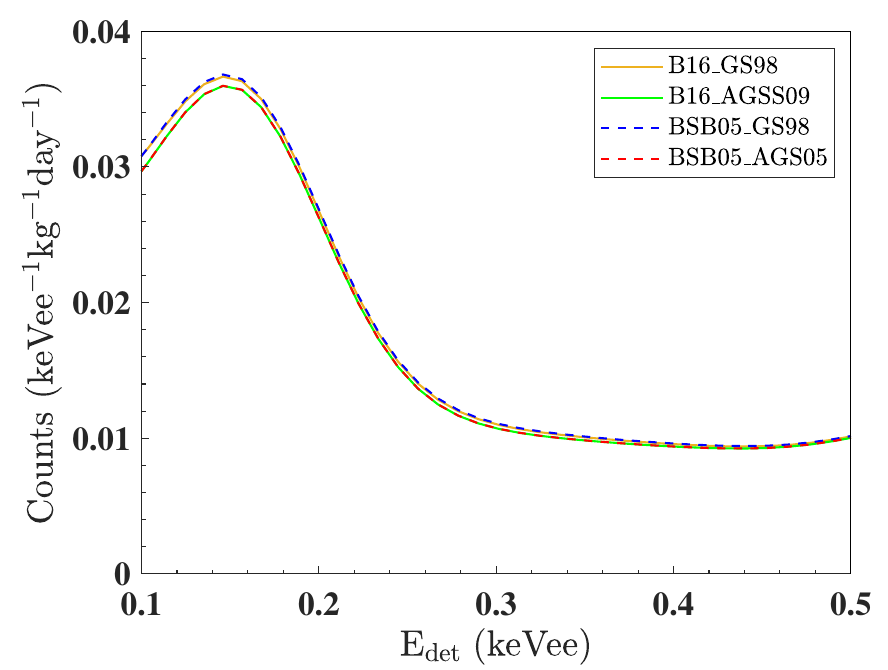}
	\caption{\textbf{Top:} Expected spectra in CDEX-50 detectors of solar neutrino with various solar models. \textbf{Bottom:} Spectra in 0.1--0.5 keVee with various solar models. The energy regions are taken for displaying the variety.}
	\label{fig:variety}
\end{figure}

In addition to the B16-GS98 solar model, there are several other solar models with different solar neutrino fluxes and consequently different background contribution. The background contributions from solar neutrino with B16-AGSS09~\cite{solar_neutrino_flux}, BSB05-GS98~\cite{solar_neutrino_flux_05}, and BSB05-AGS05~\cite{solar_neutrino_flux_05} are studied, and the variety of those contributions are shown in figure~\ref{fig:variety}. The impact of different solar models on the background level is minimal, the maximum difference between the spectra is less than 3\%. In consequence, the WIMP projected sensitivity is robust to various solar models.

\subsection{CDEX-50 background model}~\label{sec:model}
\begin{figure}[!htbp]
	\centering\includegraphics[width=\columnwidth]{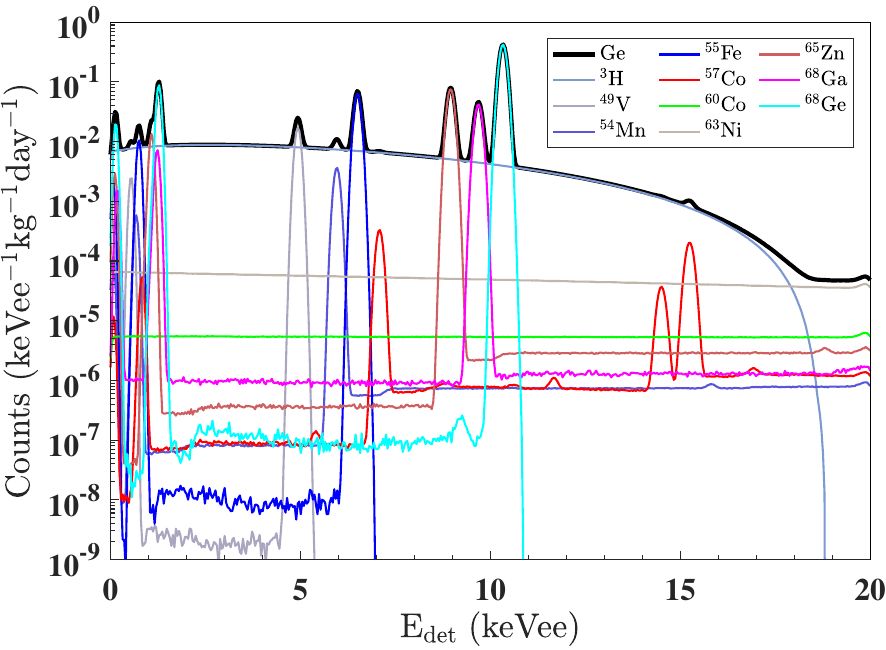}
	\centering\includegraphics[width=\columnwidth]{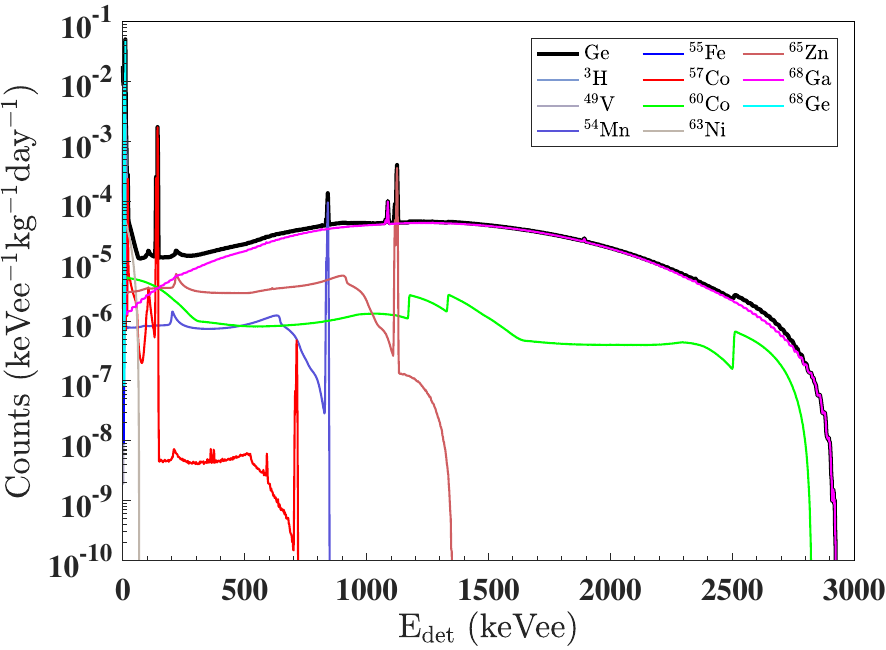}
	\caption{\textbf{Top:} Spectrum of cosmogenic radionuclides in crystal in 0--20 keVee, which are broken down into radionuclides. \textbf{Bottom:} Spectrum in 0--3000 keVee. The energy resolution is considered in both spectra.}
	\label{fig:Gespectrum}
\end{figure}

The CDEX-50 background model is built based on the background sources analysis. The spectrum from cosmogenic radionuclides in the crystal is shown in figure~\ref{fig:Gespectrum}. The $\beta^{-}$ decay from $^{3}$H dominates the contribution below the endpoint energy of 18.6 keV. Moreover, $^{63}$Ni and $^{60}$Co have $\beta^{-}$ decay with the endpoint energies of 66.9 keV and 317.9 (branching ratio of 99.88\%)/1490.3 keV, respectively. Meanwhile, their specific activities are much lower than that of $^{3}$H, leading to much lower platforms. Other radionuclides, with EC or $\beta^{+}$ decay, will release characteristic x-rays in this region, resulting in Gaussian full-energy peaks in the spectrum. In the energy region above 20 keVee, the background is mainly from cascade radiation of $\gamma$ and electrons with continuous energy from $\beta$ and EC decay. The Gaussian full-energy $\gamma$ peaks from $^{65}$Zn, $^{68}$Ga, $^{55}$Fe, and $^{57}$Co are visible in this region.

\begin{figure}[!htbp]
	\centering\includegraphics[width=\columnwidth]{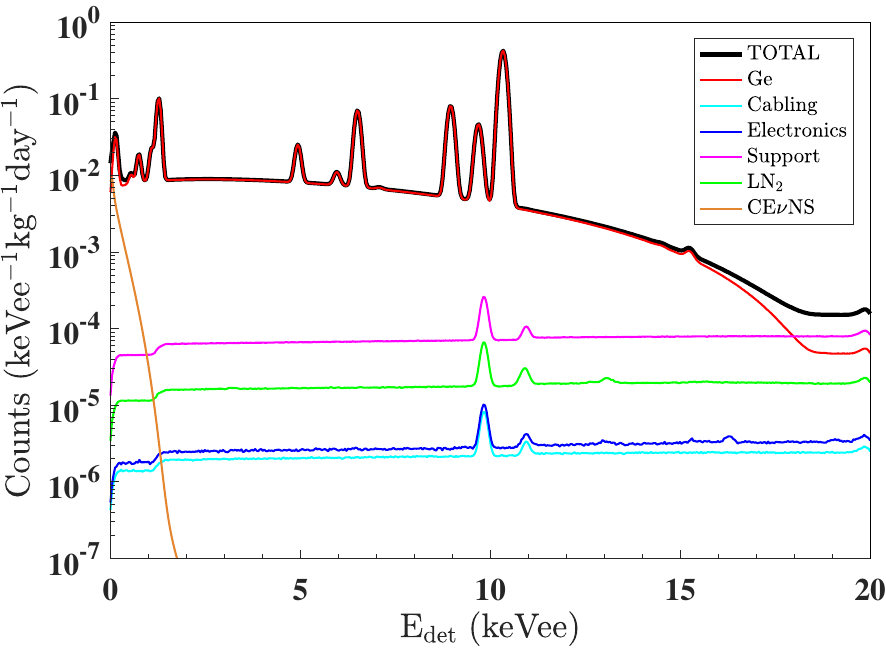}
	\centering\includegraphics[width=\columnwidth]{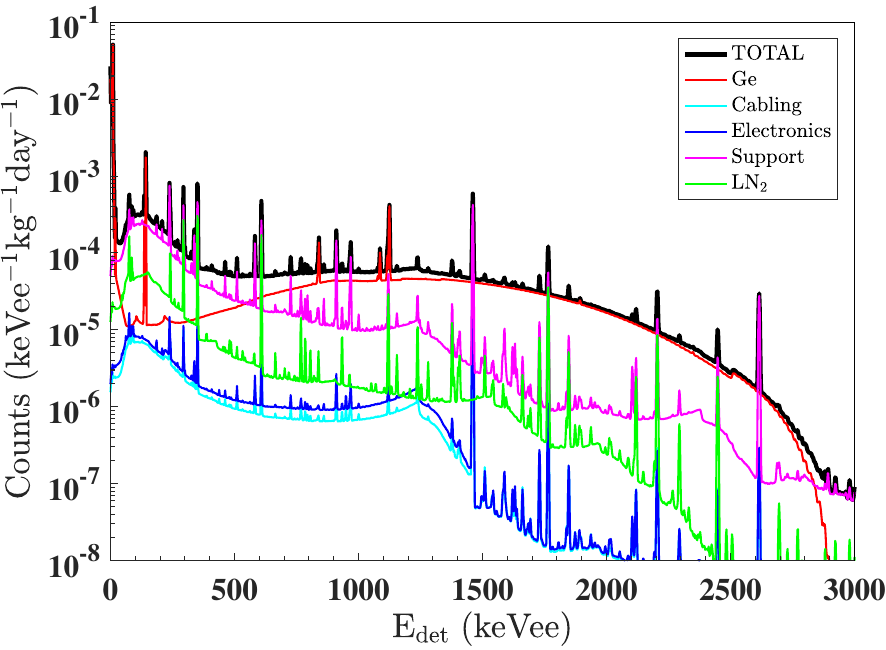}
	\caption{\textbf{Top:} Total spectrum in 0--20 keVee, which are broken down into components. \textbf{Bottom:} Spectrum in 0--3000 keVee. The contribution from CE$\nu$NS is not displayed because the effect exists in relatively low-energy regions. The energy resolution is considered in both spectra.}
	\label{fig:totalspectrum}
\end{figure}

\begin{figure}[!htbp]
	\centering\includegraphics[width=\columnwidth]{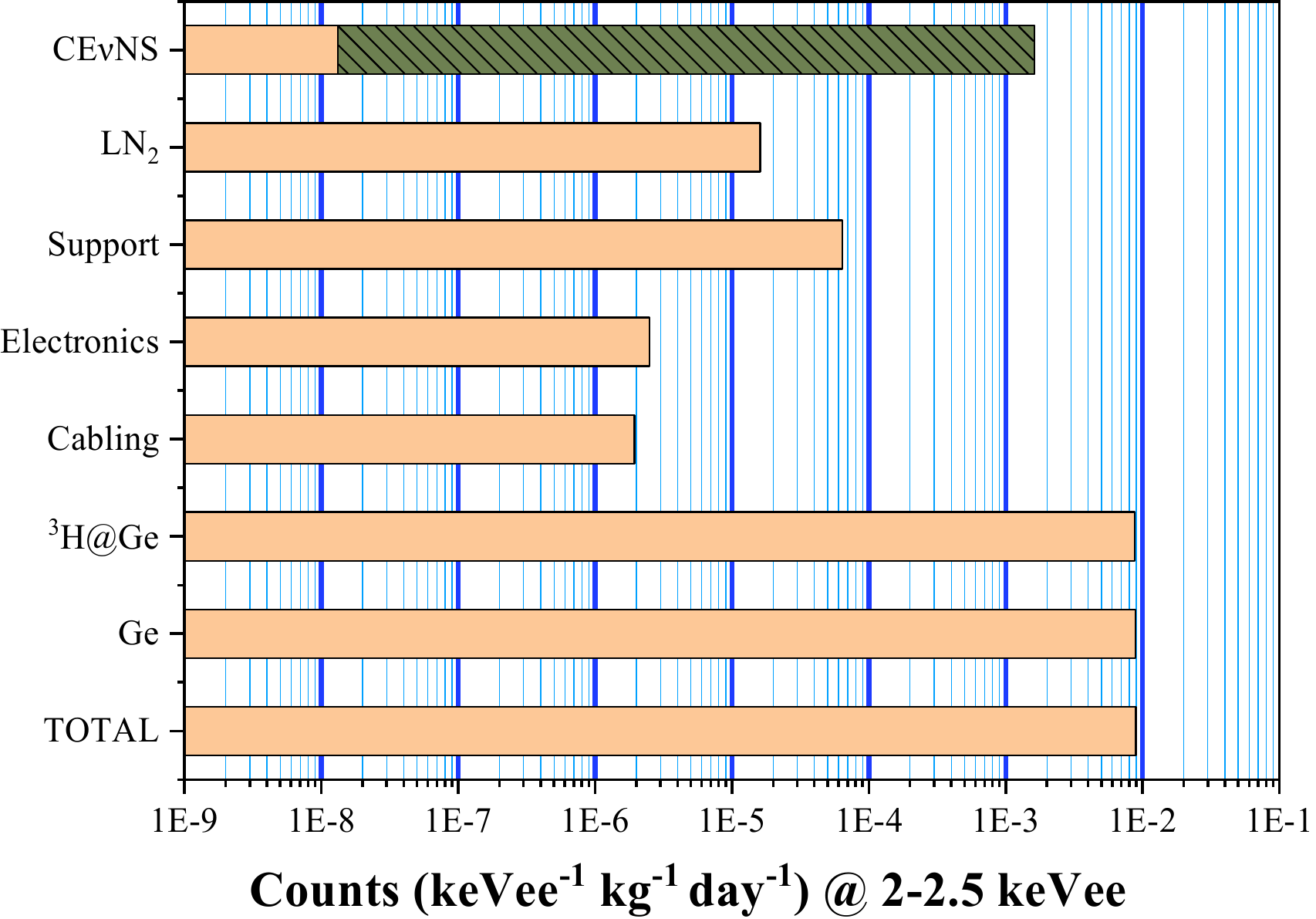}
	\caption{Contribution in ROI of various sources in CDEX-50, more than 90\% background is from $^{3}$H. The background level of CDEX-50 in ROI is $\sim$0.01 cpkkd. The green column with strips is the contribution of CE$\nu$NS in 0.16--0.5 keVee (see section~\ref{sec:cevns}).}
	\label{fig:background_level}
\end{figure}

The total spectrum and ROI contribution are shown in figure~\ref{fig:totalspectrum} and figure~\ref{fig:background_level}, respectively. Their components are categorized following table~\ref{tab:unit}. $^{3}$H dominates in the low-energy region below 18.6 keVee, whereas $^{68}$Ga and $^{60}$Co dominates in the high-energy region above $\sim$1.5 MeVee. Radioactive impurities in the Silicon-Base (classified as Support) also contribute to the background because of their relatively high specific activities, compounded by the relatively large mass and proximity of the Silicon-Base to the crystal. As the estimation of $^{222}$Rn concentration in LN$_{2}$ after purification in section~\ref{sec:LN2}, the contribution accounts for only $\sim$0.18\% of the total background in ROI. In this way, the concentration of $^{222}$Rn in LN$_{2}$ can be relaxed to $\sim$10 $\mu$Bq/kg level without major effect to the total background, lowering the criterion for purification.

\subsection{Expected spectrum in low-energy regions}~\label{sec:0-4}
\begin{figure}[!htbp]
	\centering\includegraphics[width=\columnwidth]{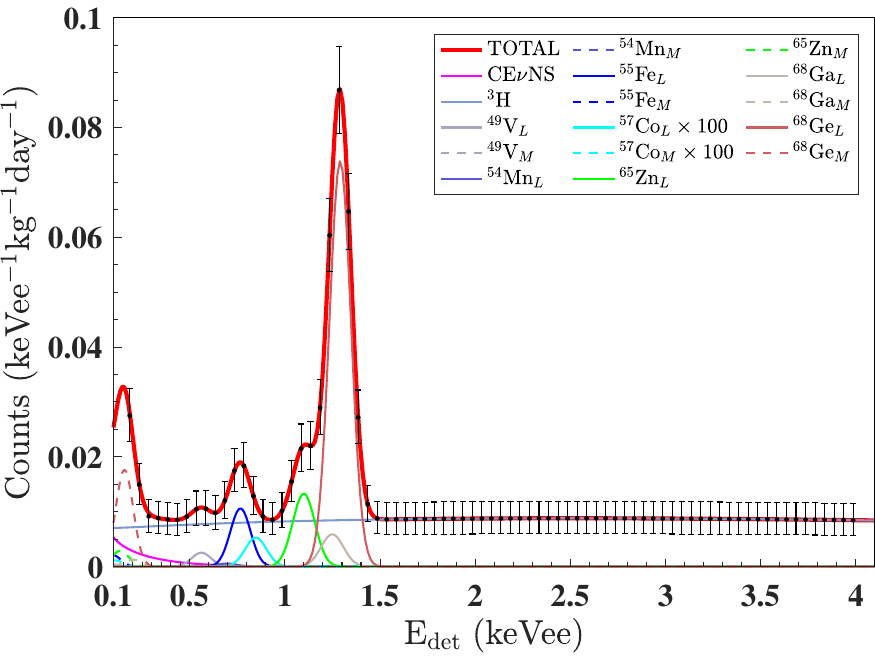}
	\caption{Expected spectrum in the low-energy region of 0.16--4.01 keVee with statistical and systematic uncertainty; the bin width is 50 eVee. The background model is the red line, which is broken down into contributors. The contribution from $^{57}$Co is multiplied by 100 for display.}
	\label{fig:0-4}
\end{figure}

The expected spectrum in low-energy region of 0.16--4.01 keVee is used to investigate the light WIMP. The statistical and systematic uncertainty are considered, the former is derived from the background model with 150 kg$\cdot$year exposure following Poisson distribution, the latter before the operation is estimated to be equal to the former based on the data analysis from CDEX-10~\cite{cdex10}. In this region, the background is mainly comprised of the curve from $^{3}$H, Gaussian full-energy peaks of L-and M-shell x-rays from cosmogenic radionuclides in germanium, and CE$\nu$NS from solar neutrino. The differential event rate of $^{3}$H decay is given by 
\begin{align}
	N(E_e) = \sqrt{E_e^2 + 2E_em_{e}}(Q-E_e)^2(E_e+m_{e})f(E_e),
	\label{equ:h3decay}
\end{align}
where $E_{e}$ is the electron energy, $m_{e}$ is the electron mass, $f(E)$ is the Fermi function~\cite{Fermi} of $^{3}$H $\beta$ decay, and $Q$ is the energy released in the decay, which is 18.6 keV. The parameters expressing the background in this region are given as follows: 
\begin{align}
	B(\theta) =&\; \theta_{1} \cdot N(E_{\rm det})\nonumber\\ &+ \sum_{i}^{2-8}\theta_{i}\cdot \frac{1}{\sqrt{2\pi}\sigma}exp(-\frac{(E_{\rm det} - E)^2}{2\sigma^2})\nonumber\\ 
	&+\sum_{i}^{2-8}r_{M/L}\cdot \theta_{i}\cdot \frac{1}{\sqrt{2\pi}\sigma}exp(-\frac{(E_{\rm det} - E)^2}{2\sigma^2})\nonumber\\ 
	&+\sum_{i}^{9-10}\theta_{i}\cdot N_{T}\int_{E_{\nu}^{min}}^{\infty}\frac{d\Phi_{norm}}{dE_{\nu}}\frac{d\sigma(E_{r},E_{\nu})}{dE_r}dE_{\nu},
	\label{equ:expression}
\end{align}
where $E_{\rm det}$ is the deposited energy, $\theta_{1}$ denotes the intensity of the curve from $^{3}$H, $\theta_{2-8}$ correspond to the intensities of the full-energy peaks of $^{49}$V, $^{54}$Mn, $^{55}$Fe, $^{57}$Co, $^{65}$Zn, $^{68}$Ga, and $^{68}$Ge x-rays, respectively; $\sigma$ and $E$ denote the energy resolution in standard deviation at the corresponding x-rays energy $E$. The M-shell/L-shell x-rays are estimated from the intensity ratio $r_{M/L}$ of M-shell/L-shell to K-shell~\cite{KMLRatio1,KMLRatio2}, $\theta_{9-10}$ denote the fluxes of $\rm ^{8}$B and $\rm hep$ solar neutrino, and $\frac{d\Phi_{norm}}{dE_{\nu}}$ is the normalized differential flux of neutrinos. The spectrum and background model based on eq.~(\ref{equ:expression}) and evaluated $\theta$ parameters are shown in figure~\ref{fig:0-4}.

\section{Projected WIMP sensitivity}\label{sec:4}
The projected sensitivity band of 90\% CL upper limits for spin-independent (SI) WIMP-nucleon couplings of CDEX-50 is estimated using the profile likelihood ratio method~\cite{PLR} for the ultra-low background level of CDEX-50. 

\subsection{WIMP signal model}
\begin{figure}[!htbp]
	\centering\includegraphics[width=\columnwidth]{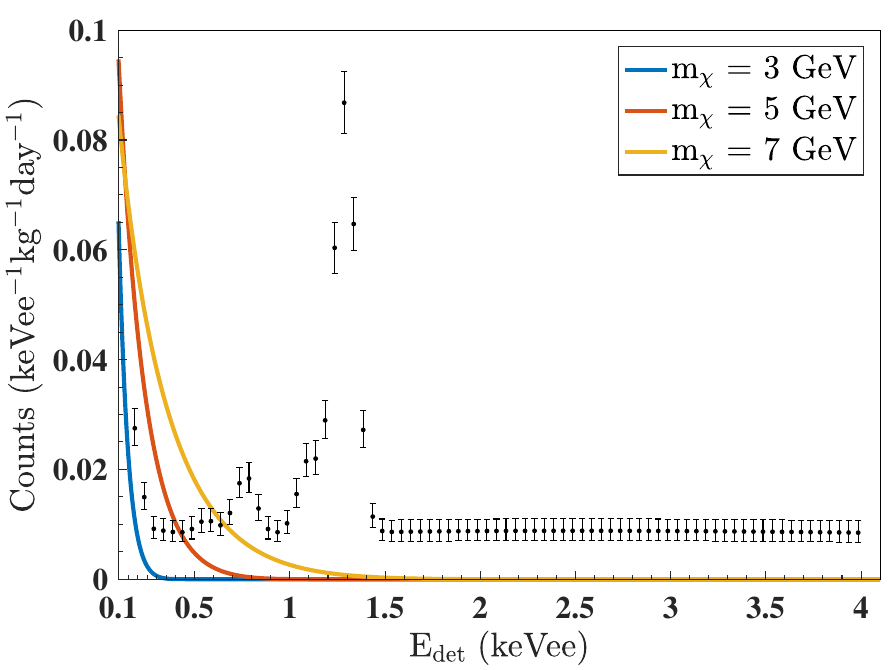}
	\caption{Expected spectra of the SI WIMP-nucleon signals for m$_\chi$ = 3, 5, 7 GeV/c$^2$ with a SI WIMP-nucleon cross section $\sigma^{\rm {SI}}_{\chi N}$ = 10$^{-43}$ cm$^2$ with considering quenching factor and energy resolution, along with the expected spectrum in the low-energy regions of 0.16--4.01 keVee.}	
	\label{fig:wimp_exp}
\end{figure}

In this work, the standard halo model (SHM)~\cite{darkmatter1,darkmatter2,darkmatter3} is adopted with a local DM density of $\rho_{\chi} = $ 0.3 GeV/cm$^3$, local standard of rest velocity of $v_{0} = $ 238 km/s, Galactic escape velocity of $v_{esc} = $ 544 km/s, and velocity of Earth with respect to the Galactic rest reference frame of $v_{E}$ = 250 km/s. For the SI WIMP-nucleon coupling, the cross-section is
\begin{equation}
	\sigma^{SI} = \frac{\mu_{N}}{\mu_{p}}^2 A^2 F^2 \sigma_{\chi N}^{SI},
\end{equation}
where $\mu_{N}$ is the reduced mass of WIMP and the target nucleus, and $\mu_{p}$ is the reduced mass of WIMP and the neutron. The differential event rate of the nuclear recoil can be written as
\begin{equation}
	\frac{d R}{d E_r} = N_{T}\frac{\rho_{\chi}}{m_\chi} \int_{v_{\rm min}} v f(\vec v, \vec v_E) d^{3}\vec v \frac{d\sigma^{SI}}{d E_r}, 
\end{equation}
where $m_\chi$ is the WIMP mass, $f(\vec v, \vec v_E)$ is the WIMP velocity distribution around Earth within the SHM, where $\vec v$ is the WIMP velocity with respect to Earth~\cite{darkmatter2} reference frame, and the $v_{\rm min}$ is the minimum velocity required to generate recoil energy $E_{r}$. The expected spectra for the SI WIMP-nucleon interaction are shown in figure~\ref{fig:wimp_exp}.

\subsection{Profile likelihood ratio method}
A statistical model based on binned likelihood, considering both the statistical and systematic uncertainty of the background model in section~\ref{sec:0-4}, is given as follows:
\begin{align}
	\mathcal{L}(m_\chi,\sigma_\chi,\theta) =& \prod_{j}[\frac{1}{{D_j}!}(S_j(m_\chi,\sigma_\chi) + B_j(\theta))^{D_j}\nonumber\\
	&\times exp(-(S_j(m_\chi,\sigma_\chi) + B_j(\theta)))\nonumber\\
	&\times \frac{1}{\sqrt{2\pi}\sigma_{syst,j}}exp(-\frac{(D_j - (S_j + B_j))^{2}}{2\sigma_{syst,j}^{2}})],
\end{align}
where $\sigma_\chi$ is the WIMP-nucleon cross-section, $S_j(m_\chi,\sigma_\chi)$ is the WIMP model count in $j$-th bin corresponding to certain $m_\chi,\sigma_\chi$; $B_j$ is the background model count in $j$-th bin corresponding to certain $\theta$ described in section~\ref{sec:0-4}, $D_j$ is the count in $j$-th bin of one certain sample generated by MC simulation, and $\sigma_{syst,j}$ is the systematic uncertainty for $j$-th energy bin. 

The profile likelihood ratio for one certain WIMP mass, used as the test statistic to test the WIMP-signal and background-only hypothesis, is given by
\begin{align}
	q(\sigma_\chi) \equiv -2log\frac{\mathcal{L}(\sigma_\chi,\hat{\hat{\theta}})}{\mathcal{L}(\hat{\sigma_\chi},\hat{\theta})},
\end{align}
where $\mathcal{L}$ is the likelihood function, $\hat{\sigma_\chi}$ and $\hat{\theta}$ are the signal and background (nuisance) parameters that maximize the $\mathcal{L}$ globally, $\hat{\hat{\theta}}$ is the background (nuisance) parameters that maximize the $\mathcal{L}$ at given $\sigma_\chi$. The $q(\sigma_\chi)$ distribution is obtained by generating $\mathcal{O}(10^{4})$ samples following the Poisson distribution with an expectation value based on the signal-plus-background model.

90\% CL upper limits distribution, from which the sensitivity band is derived, is computed using $\mathcal{O}(10^{4})$ background samples generated following the Poisson distribution with an expectation value based on the background model. 90\% CL upper limit for one certain background sample is obtained by finding the specific $\sigma_\chi$ states the test statistic of the sample is the 90\% quantile of the $q(\sigma_\chi)$ distribution. 

\begin{figure}[!tbp]
	\centering\includegraphics[width=\columnwidth]{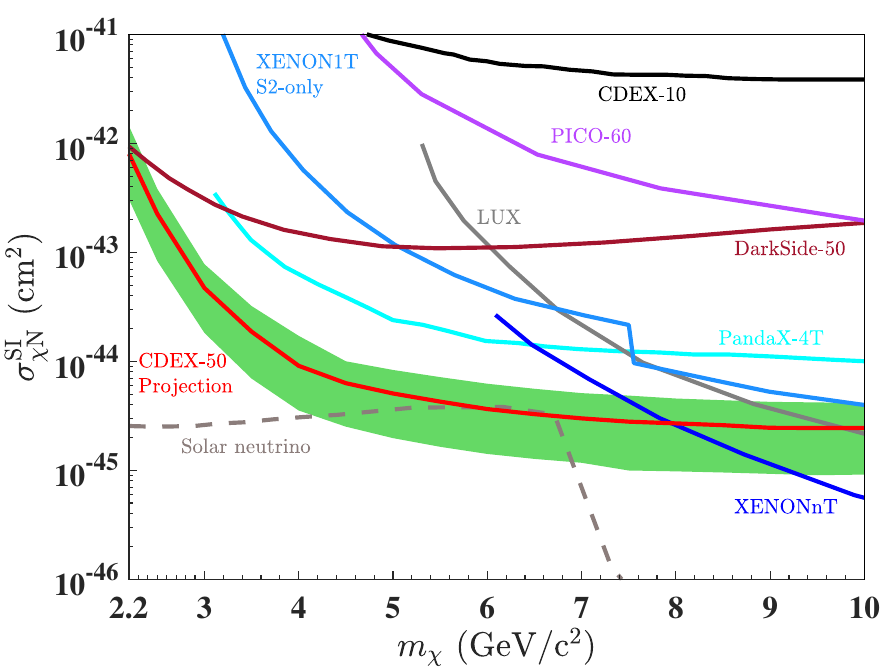}
	\caption{Projected sensitivity of CDEX-50 as 90\% CL exclusion limit (red line) on the SI WIMP-nucleon couplings with $\pm 1\sigma$ band (green). Other selected constraints from CDEX-10~\cite{cdex10}, Darkside-50~\cite{darkside}, PICO-60~\cite{PICO}, LUX~\cite{LUX}, XENON1T S2-only~\cite{xenon1t}, XENONnT~\cite{xenonnt}, and PandaX-4T~\cite{PandaX-4T} are superimposed. The gray dashed line represents the neutrino discovery limit for Ge-based experiments~\cite{neutrino_floor}.}
	\label{fig:wimp}
\end{figure}

The expected CDEX-50 sensitivity on the SI WIMP-nucleon couplings is shown in figure~\ref{fig:wimp}. With an exposure objective of 150 kg$\cdot$year and analysis threshold of 160 eVee, the expected sensitivity on the SI WIMP-nucleon couplings is estimated to reach a cross-section of 5.1 $\times$ 10$^{-45}$ cm$^{2}$ for a WIMP mass of 5 GeV/c$^{2}$ at 90\% CL. This science goal will correspond to the most sensitive results for WIMPs with a mass of 2.2--8 GeV/c$^{2}$. 

\section{Summary}
In this paper, we present a comprehensive overview of the CDEX-50 detector unit components and the geometry of the CDEX-50 detector array. We analyze the background sources arising from the environment, detector components, and solar neutrino using radioassay and MC simulation. A background model is built by incorporating the Geant4 simulation of radionuclides and theoretical evaluation of CE$\nu$NS from solar neutrino. By applying self-anticoincidence measures among the array, the background level in ROI is estimated to be $\sim$0.01 cpkkd. We project the CDEX-50 sensitivity on the SI WIMP-nucleon couplings with a 150 kg$\cdot$year exposure and 160 eVee analysis threshold using the profile likelihood ratio method and background model. As a result, a cross-section of 5.1 $\times$ 10$^{-45}$ cm$^{2}$ for $m_\chi$ = 5 GeV/c$^{2}$ at 90\% CL is obtained. This outcome provides the most sensitive result for $m_\chi$ in 2.2--8 GeV/c$^{2}$ to date. The result is more than three orders of magnitude beyond the CDEX-10 experiment for a larger exposure and a much lower background level, which can be attributed to the improved purity of detector components, stringently controlled germanium exposure, and LN$_{2}$ tank shielding.

The CDEX-50 experimental sensitivity on WIMP is highly dependent on the background in the low-energy region where the L- and M-shell x-rays originate from cosmogenic radionuclides contributes several Gaussian peaks as shown in figure~\ref{fig:0-4}. These peaks play a key role in forming the spectra whereas their energies are close, so understanding of them is significant to obtain competitive experimental sensitivity. To identify the L- and M-shell x-rays, the corresponding K-shell x-rays in $\mathcal{O}$(1) keVee which are relatively distinguishable can impose strong constraints on L- and M-shell x-rays~\cite{cdex_darkpthoton,cdex10_exotic}. Moreover, the calibration of energy resolution of the detectors in the low-energy region is vital to obtaining the accurate information of the K-shell x-rays and the spectra including L- and M-shell x-rays. 

\begin{figure}[!tbp]
	\centering\includegraphics[width=\columnwidth]{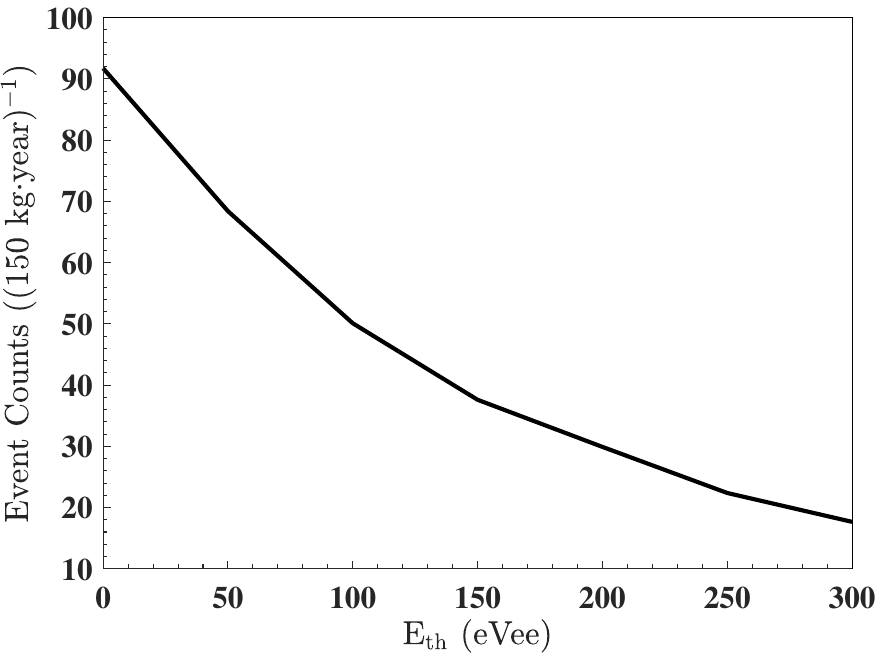}
	\caption{Expected observed events with deposited energy above the threshold of CDEX-50 detector array for a 150 kg$\cdot$year exposure with various energy thresholds, which is $\sim$36 for the expected energy threshold of 160 eVee.}
	\label{fig:CEvNS}
\end{figure}

As mentioned in section~\ref{sec:cevns}, CE$\nu$NS from solar neutrino can be detected by CDEX-50 located in CJPL. The detection of solar neutrinos~\cite{solarneutrino} is significant in the study of neutrinos and the structure of the Sun. However, it is challenging because of the low expected event rate above the threshold of most detectors. The relation between the threshold and the expected observed events of CE$\nu$NS from solar neutrino for germanium detectors featuring low energy threshold is shown in figure~\ref{fig:CEvNS}. 
From previous analysis in section~\ref{sec:cevns}, signals from CE$\nu$NS is remarkable in 0.16--0.5 keVee. For a CE$\nu$NS discovery evaluation of CDEX-50, the statistical significance of the CE$\nu$NS signal in this range is estimated to be 29/$\sqrt{210}$ $\approx$ 2.0$\sigma$, where 29 and 210 are the CE$\nu$NS count and the background count excluding CE$\nu$NS in this energy region, respectively. This significance can be improved by extending the exposure, and lowering the background level or analysis threshold in the future. As the background level of CDEX-50, the platform from CE$\nu$NS below 0.25 keVee is as significant as $^{3}$H while in this range the Gaussian peaks from M-shell x-rays dominates, the platform from $^{3}$H becomes the main concern from 0.25--0.5 keVee where the contribution from CE$\nu$NS ends. However, careful study of CE$\nu$NS is need for a lower background in the future.

\begin{acknowledgments}
This work was supported by the National Key Research and Development Program of China (Grants No. 2017YFA0402200, No. 2022YFA1605000) and the National Natural Science Foundation of China (Grants No. 12322511, No. 12175112, No. 12005111, and No. 11725522). We would like to thank CJPL and its staff for hosting and supporting the CDEX project. CJPL is jointly operated by Tsinghua University and Yalong River Hydropower Development Company. We acknowledge the Center of High performance computing, Tsinghua University for providing the facility support. 
\end{acknowledgments}

\bibliography{CDEX50bkg.bib}

\end{document}